\documentclass[a4paper]{spie}  

\usepackage{amsmath,amsfonts,amssymb}
\usepackage{graphicx}
\usepackage[colorlinks=true, allcolors=blue]{hyperref}
\usepackage{float}
\usepackage{pifont}

\title{The VIS detector system of SOXS}

\author[a,c]{Rosario Cosentino} 
\author[b]{Matteo Aliverti} 
\author[c]{Salvatore Scuderi}     
\author[b]{Sergio Campana} 
\author[d]{Riccardo Claudi}  
\author[e]{Pietro Schipani}
\author[d]{Andrea Baruffolo}   
\author[i,f]{Sagi Ben-Ami}   
\author[g]{L. H. Mehrgan}
\author[g]{Derek Ives}
\author[d]{Federico Biondi} 
\author[g,m]{Anna Brucalassi} 
\author[e]{Giulio Capasso}   
\author[h]{Francesco D'Alessio}  
\author[b]{Paolo D'Avanzo} 
\author[i]{Oz Diner} 
\author[s,u]{Hanindyo Kuncarayakti} 
\author[c]{Matteo Munari}
\author[i]{Adam Rubin}
\author[h]{Fabrizio Vitali} 
\author[l]{Jani Achr\'{e}n}
\author[m]{Jos\'{e} Antonio Araiza-Duran}
\author[n]{Iair Arcavi}  
\author[b]{Andrea Bianco} 
\author[d]{Enrico Cappellaro}
\author[e]{Mirko Colapietro} 
\author[e]{Massimo Della Valle} 
\author[e]{Sergio D'Orsi}
\author[d]{Daniela Fantinel}
\author[o]{Johan Fynbo}
\author[p]{Avishay Gal-Yam}
\author[b]{Matteo Genoni}
\author[q]{Mika Hirvonen}
\author[s,u]{Jari Kotilainen}
\author[r]{Tarun Kumar}
\author[b]{Marco Landoni}
\author[q]{Jussi Lehti} 
\author[s]{Gianluca Li Causi} 
\author[d]{Luca Marafatto} 
\author[j]{Seppo Mattila}
\author[b]{Giorgio Pariani} 
\author[m, k]{Giuliano Pignata}
\author[i]{Michael Rappaport} 
\author[d]{Davide Ricci}
\author[b]{Marco Riva}
\author[d]{Bernardo Salasnich}
\author[c]{R. Zanmar Sanchez}
\author[t]{Stephen Smartt} 
\author[d]{Massimo Turatto}

\affil[a]{ INAF -- Fundaci\'{o}n Galileo Galilei, -- Bre\~{n}a Baja, Spain }
\affil[b]{INAF - Osservatorio Astronomico di Brera, Merate, Italy }
\affil[c]{INAF - Osservatorio Astrofisico di Catania, Catania, Italy }
\affil[d]{INAF - Osservatorio Astronomico di Padova, Padua, Italy }
\affil[e]{INAF - Osservatorio Astronomico di Capodimonte, Naples, Italy }            
\affil[f]{Harvard-Smithsonian Center for Astrophysics, Cambridge, USA }      
\affil[g]{ESO - European Southern Observatory, Garching, Germany }                   
\affil[h]{INAF - Osservatorio Astronomico di Roma, Rome, Italy }               
\affil[i]{Weizmann Institute of Science, Rehovot, Israel }  
\affil[j]{University of Turku, Turku, Finland }          
\affil[l]{Incident Angle Oy, Turku, Finland }                
\affil[k]{Millennium Institute of Astrophysics (MAS), Santiago, Chile }                                    
\affil[m]{Universidad Andres Bello, Santiago, Chile }                          
\affil[n]{Tel Aviv University, Tel Aviv, Israel }                              
\affil[o]{Dark Cosmology Centre, Copenhagen, Denmark }                         
\affil[p]{Weizmann Institute of Science, Rehovot, Israel }                     
\affil[q]{ASRO - Aboa Space Research Oy, Turku, Finland }                      
\affil[r]{University of Turku, Turku, Finland }                                
\affil[s]{INAF - Istituto di Astrofisica e Planetologia Spaziali, Rome, Italy} 
\affil[t]{Queen's University Belfast, Belfast, UK }     
\affil[u]{FINCA - Finnish Centre for Astronomy with ESO, Turku, Finland }      

\begin{document} 
\maketitle

\begin{abstract}
SOXS will be a unique spectroscopic facility for the ESO NTT telescope able to cover the optical and NIR bands thanks to two different arms: the UV-VIS (350-850 nm), and the NIR (800-1800 nm). In this article, we describe the design of the visible camera cryostat and the architecture of the acquisition system. The UV-VIS detector system is based on a e2v CCD 44-82, a custom detector head coupled with the ESO continuous flow cryostats (CFC) cooling system and the NGC CCD controller developed by ESO.
This paper outlines the status of the system and describes the design of the different parts that made up the UV-VIS arm and is accompanied by a series of contributions describing the SOXS design solutions (Ref. ~\citenum{soxsold,soxsaliverti,soxsbiondi,soxsbrucalassi,soxscapasso,soxsclaudi,soxssanchez,soxsschipani,soxsricci,soxsrubin,soxsvitali,ngcpaper}).

\end{abstract}

 \footnote{Author: cosentino@tng.iac.es}



\section{The Detector System}
The UV-VIS CCD Detector System for SOXS is designed to reach a wavelength response from 350 to 850 nm and consists of the following sub-systems: 
\begin{enumerate}
\setlength{\itemsep}{0pt}
\item e2v CCD44-82 2K x 4K CCD
\item CFC cooling system
\item	detector head
\item	CCD controller unit and power supply to operate the CCD detector head, with cable set to connect them
\item	shutter
\item	commercial shutter control unit with cable set
\item	temperature controller with cable set
\end{enumerate}

\subsection{VIS Detector}
\begin{figure} [H]
\begin{minipage}[c]{0.6\linewidth}
\includegraphics[width=\linewidth]{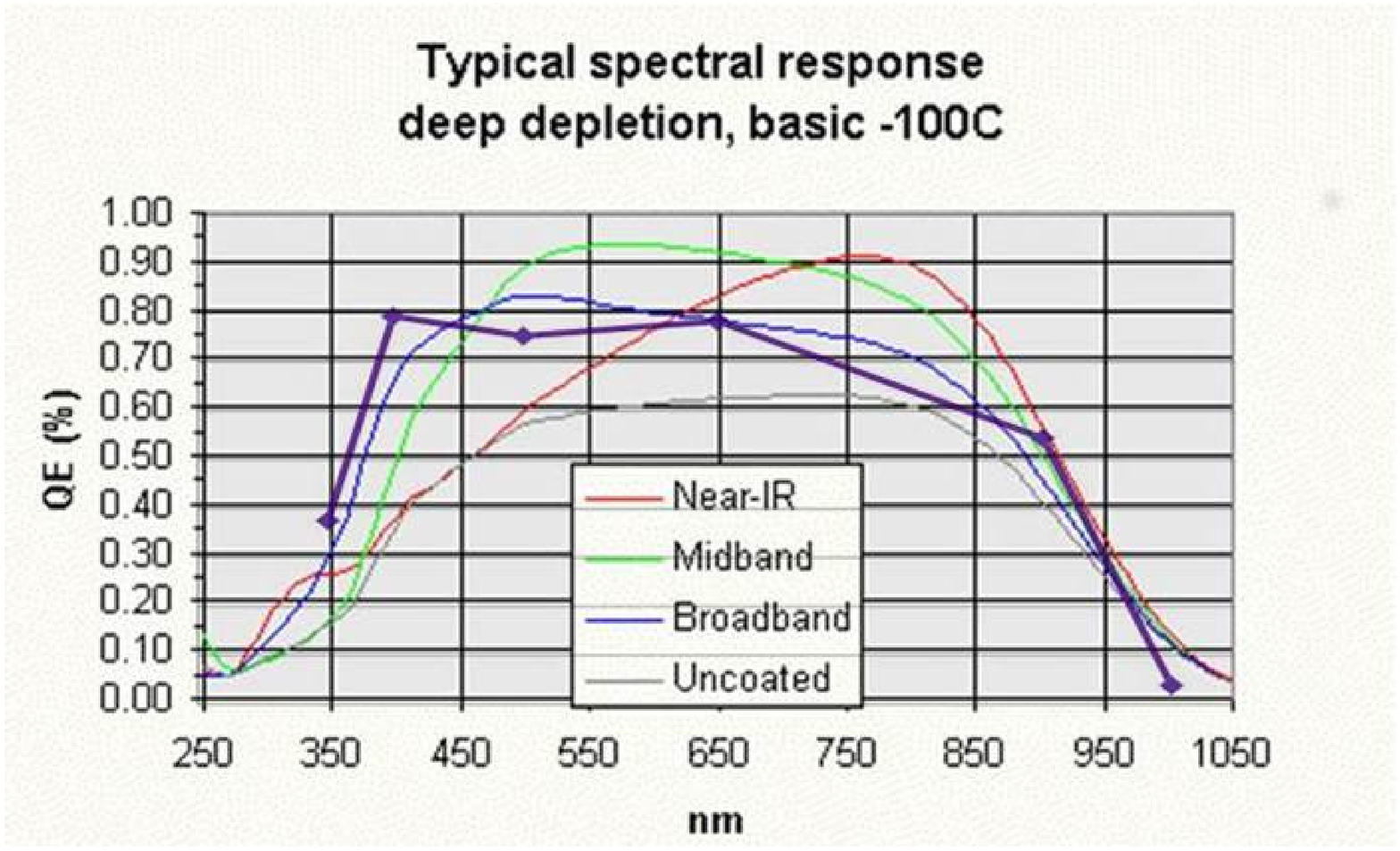}
\caption{QE vs. wavelength for different coatings as provided by the manufacturer. The selected CCD is the midband, the purple points are our specific CCD QE data, taken from the e2v technical note}
\label{fig1} 
\end{minipage}
\hfill
\begin{minipage}[c]{0.3\linewidth}
\includegraphics[width=\linewidth]{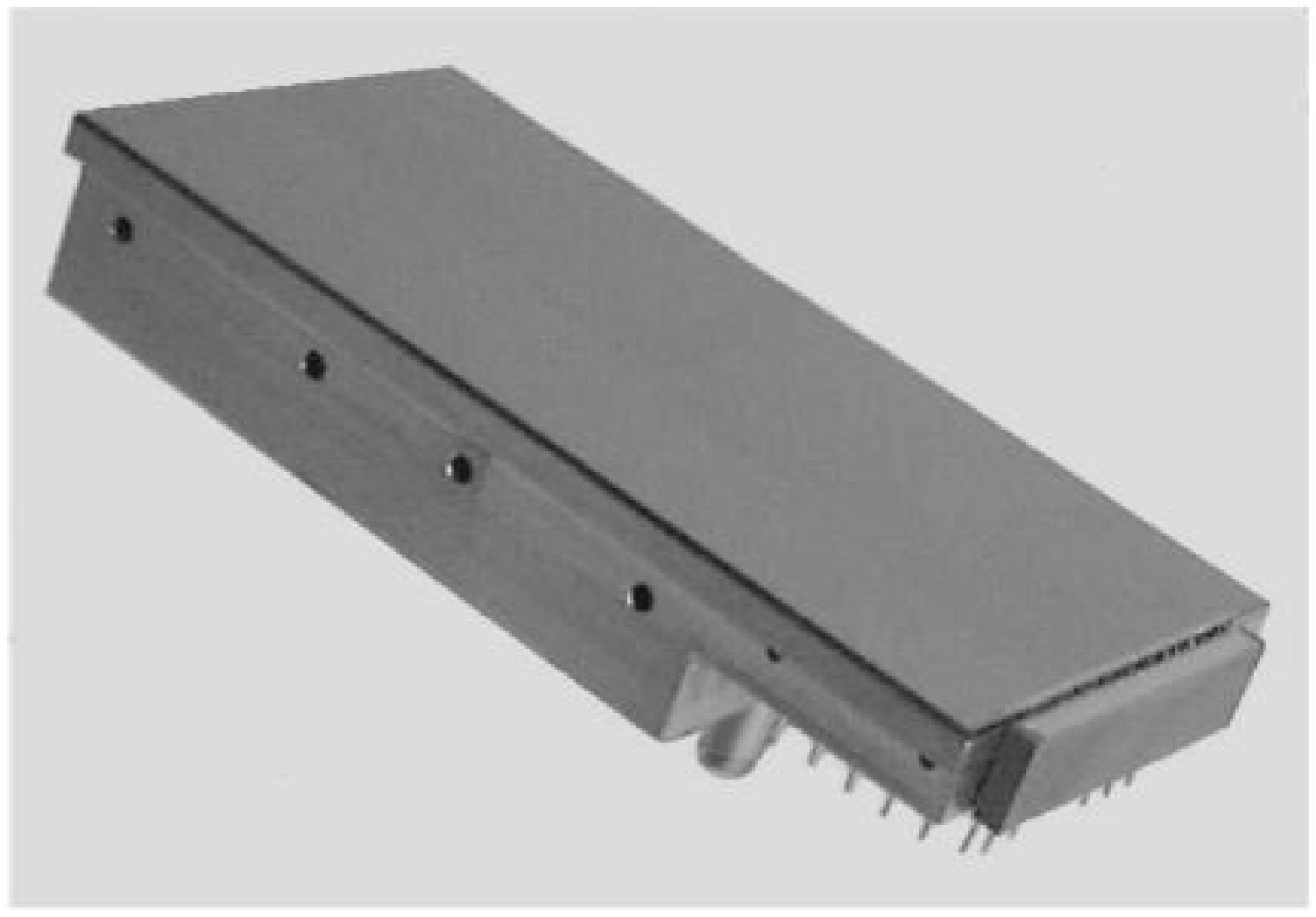}
\caption{CCD44-82 back-illuminated CCD sensor}
\label{fig2}
\end{minipage}%
\end{figure}

The detector chosen for the SOXS UV-VIS arm is an e2v CCD44-82. This detector is a high performance, back illuminated CCD with a 15.0 $\mu$m square pixel and an image area of 30.7x61.4 mm and is characterized by a high Quantum Efficiency (QE). In figure~\ref{fig1} is shown the QE of this detector for  different coatings.
This detector technology was chosen for the following reasons:
\begin{enumerate}
\setlength{\itemsep}{0pt}
\item devices manufactured with this technology meet or exceed the requirements
\item INAF has experience with devices with this technology and it is considered to be of low risk
\end{enumerate}

\begin{table}[H]
\caption{CCD main characteristics.} 
\label{tab1}
\begin{center}       
\begin{tabular}{|c|c|} 
\hline
\rule[-1ex]{0pt}{3.5ex}  Detector & CCD44-82  \\
\hline
\rule[-1ex]{0pt}{3.5ex} Chip type &	Thinned back illuminated \\
\hline
\rule[-1ex]{0pt}{3.5ex} Pixel size	& 15 $\mu$m \\
\hline
\rule[-1ex]{0pt}{3.5ex} Area (pixels)  & 2048 x 4096 \\
\hline
\rule[-1ex]{0pt}{3.5ex} Area (mm) &	30.7 x 61.4 \\
\hline
\rule[-1ex]{0pt}{3.5ex} QE at 500 nm &	90\% \\
\hline
\rule[-1ex]{0pt}{3.5ex} Coating &	yes\\
\hline
\rule[-1ex]{0pt}{3.5ex} Flatness	& Better than 20 $\mu$m peak to valley\\
\hline
\rule[-1ex]{0pt}{3.5ex} Peak signal &	200 ke-/pixel\\
\hline
\rule[-1ex]{0pt}{3.5ex} CTE Serial OSL	& 99.9993\%\\
\hline
\rule[-1ex]{0pt}{3.5ex} CTE Serial OSR	& 99.9999\%\\
\hline
\rule[-1ex]{0pt}{3.5ex} CTE Parallel	& 99.9998\%\\
\hline
\end{tabular}
\end{center}
\end{table} 

\subsection{The camera housing}

\begin{figure} [H]
   \begin{center}
   \begin{tabular}{c} 
   \includegraphics[height=5cm]{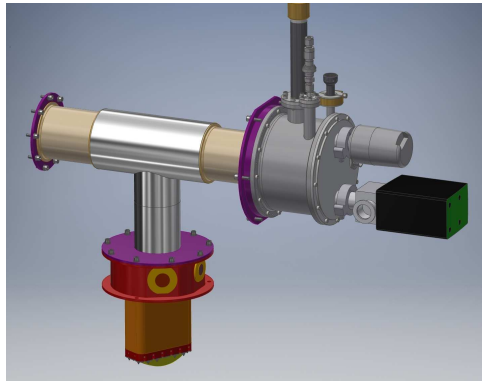}
   \end{tabular}
   \end{center}
   \caption[VIS Camera] 
   { \label{fig16} Drawing of the VIS detector system}
   \end{figure}

The detector head and wiring are based on the heritage of existing instruments at ESO. The head is coupled with a Continuous Flow Cryostat (CFC) cooling system , successfully adopted in several ESO projects. The UV-VIS camera is made up by the ESO CFC cooling system coupled through a bellows with the custom detector head that allows its placement into the UV-VIS arm of the spectrograph figure~\ref{fig16}.  Due to the reduced space available in the spectrograph, the front part of the camera, where the  light will be incoming, has a special rectangular design called 'nose'.

\subsection{Detector head and wiring}

The detector wiring, i.e. the baseplate and flex circuits, will provide the components to safely mount the CCD and to conduct signals from the CCDs to the external wall of the cryostat. It also provides the mounting and wiring for the temperature sensors and heater that stabilize the detector temperature and monitor cryostat performance. 
The detector wiring will place minimal heat load on the cooling system. It should not introduce noticeable crosstalk or readout noise, and it should also be flexible enough to accommodate shrinkage during cooling down. It should not add appreciably to the outgassing and should use as much as possible previously tested cryogenic compatible materials and components.
In the design and choice of fabrication materials and techniques for the detector head wiring, the following rules have been followed:

\begin{itemize}
\setlength{\itemsep}{0pt}
\item the preamplifiers are directly attached to the cryostat or very close to it
\item the clocks-filters and wave-shaping (passive components) are placed in  a shielded box attached to the cryostat
\item clock, bias, output, and temperature signal groups are kept separate, with ground shields in between
\item the thermal load of wiring on cryogenic cooling is kept as low as possible
\end{itemize}

\subsection{CCD controller unit}

The detector controller provides the biases and the correct sequences of clocks and analog video processing to drive and acquire the images of the CCDs. 
The dead time is defined as the interval between the end of one exposure (closed shutter) to the start of the next one (open shutter) and it obviously includes the read time and also other software overheads, e.g. for data storage and shutter control. 
The CCD controller will be the NGC (New General detector Controller) system. This controller, designed by ESO to replace the former FIERA and IRACE devices, is a single multipurpose controller able to manage both CCDs and IR arrays.The NGC is the current controller for ESO instrumentation and it will be adopted in SOXS project as well to comply with ESO standards and to reduce any compatibility failure with ESO hardware and software systems.
The NGC is based on newest hardware standard (FPGA), it has the same hardware core for both optical and IR controllers and, in addition, is full compatible with ESO software  and it includes all the good features of previous systems.

The NGC consists of five major components:
\begin{itemize}
\setlength{\itemsep}{0pt}
\item	Data Acquisition Computer with integrated PCI interface (DAQC)
\item	Detector Front-End (DFE)
\item	Detector Cryostat Cables (DCC)
\item	Detector Front-end Power Supply (DFPS)
\item	Detector Preamplifier (DPA)
\end{itemize}
The main NGC features are:
\begin{itemize}
\setlength{\itemsep}{0pt}
\item	high speed link
\item	the core element on each board is a Xilinx FPGA
\item	digital parts and the sequencer are implemented in the FPGA
\item	compact system
\item	full compatibility with ESO software standard 
\end{itemize}

The controller has to be located near the detector, because the maximum cable length must be 2 m.

The control system is composed by: 
\begin{enumerate}
\setlength{\itemsep}{0pt}
\item	LLCU, which is a Linux machine (DELL or an industrial PC)
\item	PCI-Express card
\item	front-end box (preamplifier)
\item	power supply, including the cable to the NGC
\item	two NGC modules for operations with CCD detectors
\item	fiber interfaces, including fiber cable
\item	CCD temperature control
\item	shutter driver
\end{enumerate}

The NGC system has three possible housings:
compact, fan-less and water-cooling. The housing is selected according to the setup where the controller is located. In SOXS we need to locate the controller close to the instrument because the cable between the cryostat and the device must be shorter than 2 m. Therefore, we selected the water-cooled housing to prevent any thermal dissipation into the telescope Nasmyth room.

\begin{figure}
\begin{minipage}[c]{0.2\linewidth}
\includegraphics[width=\linewidth]{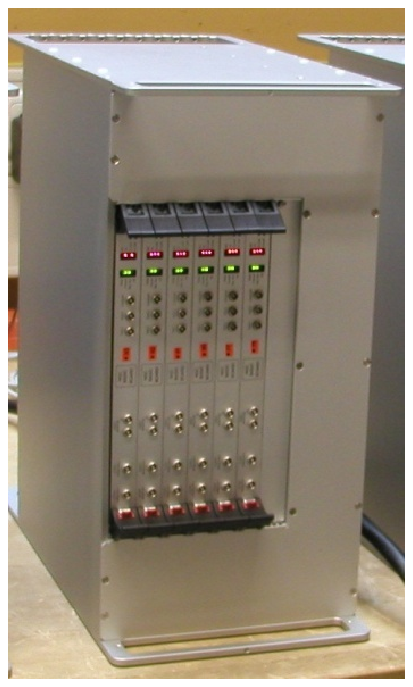}
\caption{The NGC water-cooled housing}
\label{fig3} 
\end{minipage}
\hfill
\begin{minipage}[c]{0.75\linewidth}
\includegraphics[width=\linewidth]{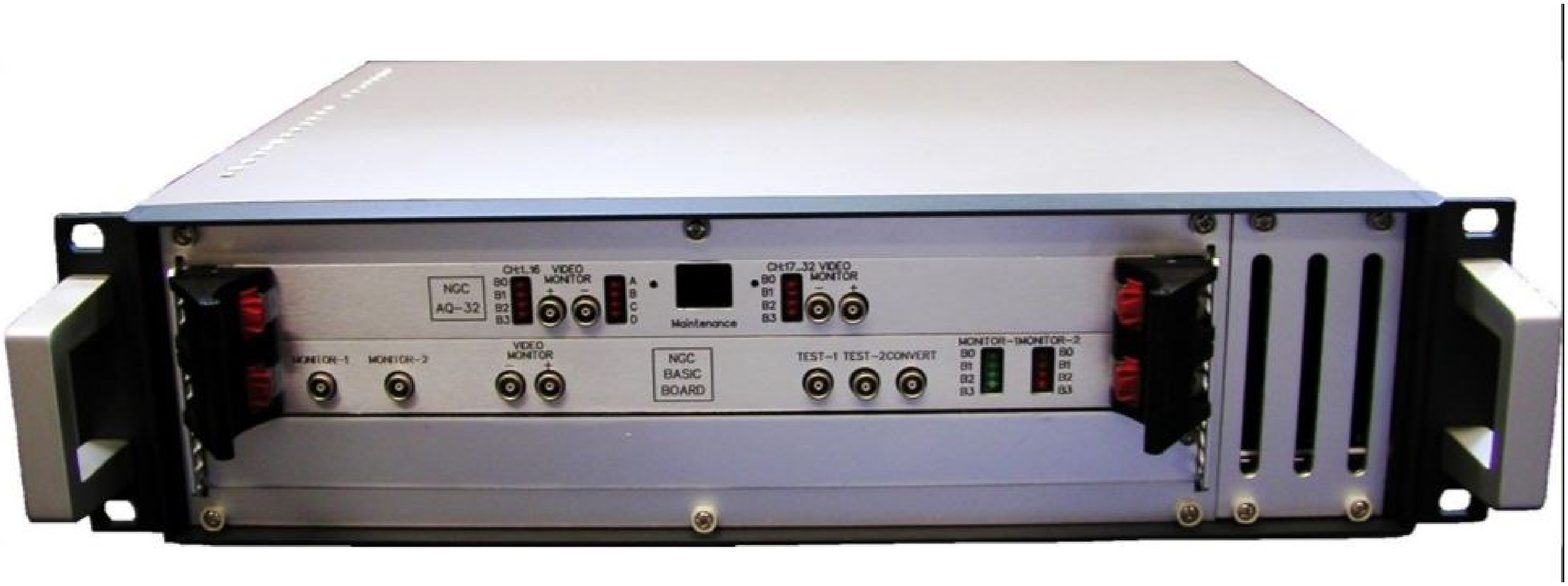}
\caption{The NGC Two-Slot system housing}
\label{fig4}
\end{minipage}%
\end{figure}

The power supply for the NGC device is a 19-inch 3HE rack-mountable unit, which may be at a distance from the detector Front End (DFE) of up to 12 meters (see figure~\ref{fig9}).

   \begin{figure} [H]
   \begin{center}
   \begin{tabular}{c} 
   \includegraphics[height=7cm]{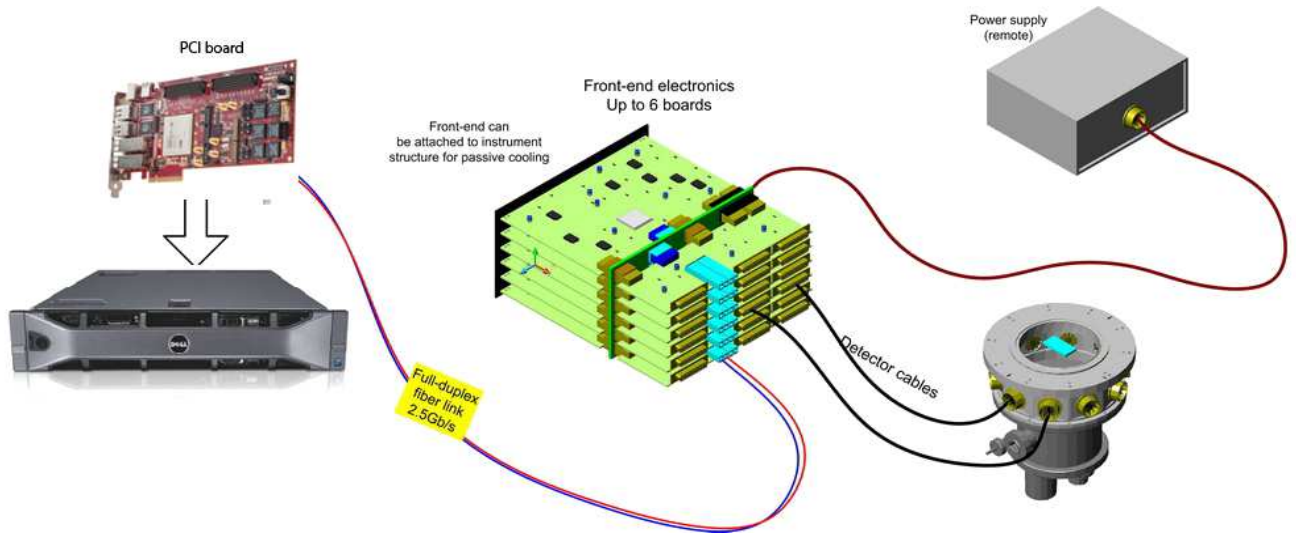}
   \end{tabular}
   \end{center}
   \caption[The New General detector Controller (NGC)] 
   { \label{fig5} The ESO New Generation Controller functional diagram.}
   \end{figure} 
   
\subsubsection{PCI interface Board (DAQC)}  
\begin{figure} [H]
   \begin{center}
   \begin{tabular}{c} 
   \includegraphics[height=7cm]{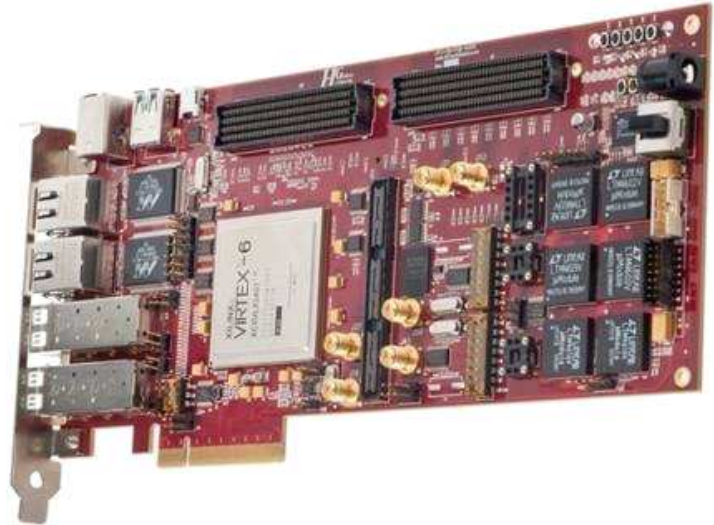}
   \end{tabular}
   \end{center}
   \caption[The PCIe interface board ] 
   { \label{fig6} The PCIe interface board.}
   \end{figure}

An ESO custom-made PCI64 board (figure~\ref{fig6}) with a fiber-optic connection to the Front End is installed in the Data Acquisition Computer (DACQ). The maximum theoretical bandwidth per interface is 256MB/s, which matches the 2.5 GBit/s fiber transmission rate. The bandwidth for actual data transmission is about 20\% lower.


\subsubsection{Detector front-end (DFE)}  
The DFE has to be located close to or on the instrument (maximum detector cable length 2m; depending on the application, an extension up to 3m may be possible).
The module, from which the VIS NGC DFE can be built, is one Basic Board that contains four video channels and clock/bias generation.
In order to prevent damage from overheating, all housings should be equipped with a thermal sensor that can shut off the power to the NGC box. 
   \begin{figure} [H]
   \begin{center}
   \begin{tabular}{c} 
   \includegraphics[height=7cm]{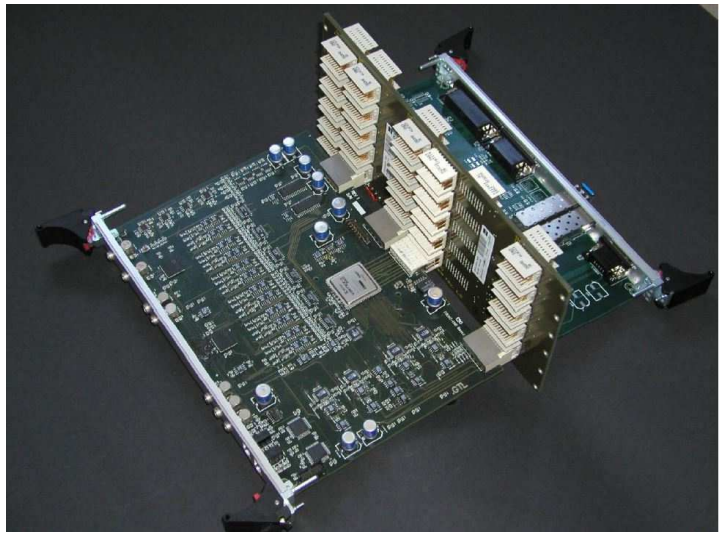}
   \end{tabular}
   \end{center}
   \caption[NGC system with Basic Board, Backplane and Transition Board (VIS arm)] 
   { \label{fig7} NGC system with Basic Board, Backplane and Transition Board (VIS arm).}
   \end{figure} 
   
\subsubsection{Detector Cryostat Cables (DCC)}  
The Detector Cryostat Cable is the connection between the  DFE and the  VIS camera (figure~\ref{fig5}). It contains the clock and bias lines and the video signals. The maximum length depends on application (2-3 m).
 
 \subsubsection{Detector Front-end Power Supply (DFPS)}  
The power supply for the DFE is a 19-inch 3HE rack-mountable unit which may be at a distance from the DFE of up to 12 meters.
   \begin{figure} [H]
   \begin{center}
   \begin{tabular}{c} 
   \includegraphics[height=4cm]{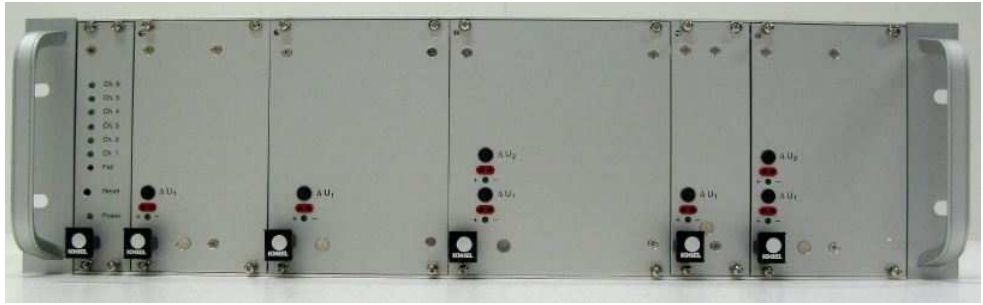}
   \end{tabular}
   \end{center}
   \caption[Power supply (front view)] 
   { \label{fig9} Power supply (front view).}
   \end{figure} 
   
\subsubsection{Detector Preamplifier (DPA)}  
\begin{figure} [H]
   \begin{center}
   \begin{tabular}{c} 
   \includegraphics[height=4cm]{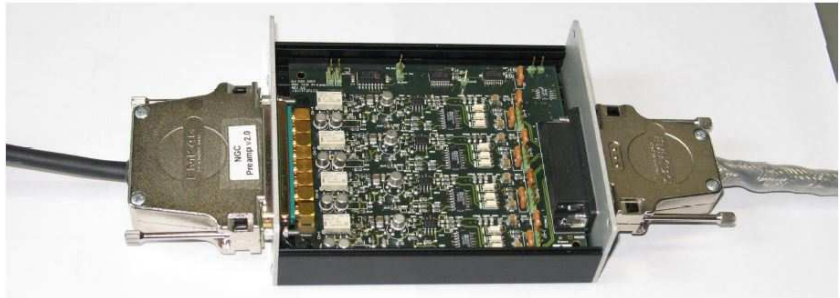}
   \end{tabular}
   \end{center}
   \caption[Detector Preamplifier (DPA)] 
   { \label{fig12} Detector Preamplifier (DPA).}
   \end{figure}

The CCD preamplifier is placed between the cryostat and the DFE.
The location of the Detector Preamplifier is application dependent; in the SOXS case of an optical detector the preamplifier is located outside the cryostat. It features 4 video channels with 16 software-selectable gains and 8 software-selectable bandwidths per channel.
Its outer dimensions are 100 x 120 x 40 mm\textsuperscript{3}. The preamp is show in figure~\ref{fig12}.

\subsection{Shutter control}
\begin{figure} [H]
\begin{minipage}[c]{0.25\linewidth}
\includegraphics[width=\linewidth]{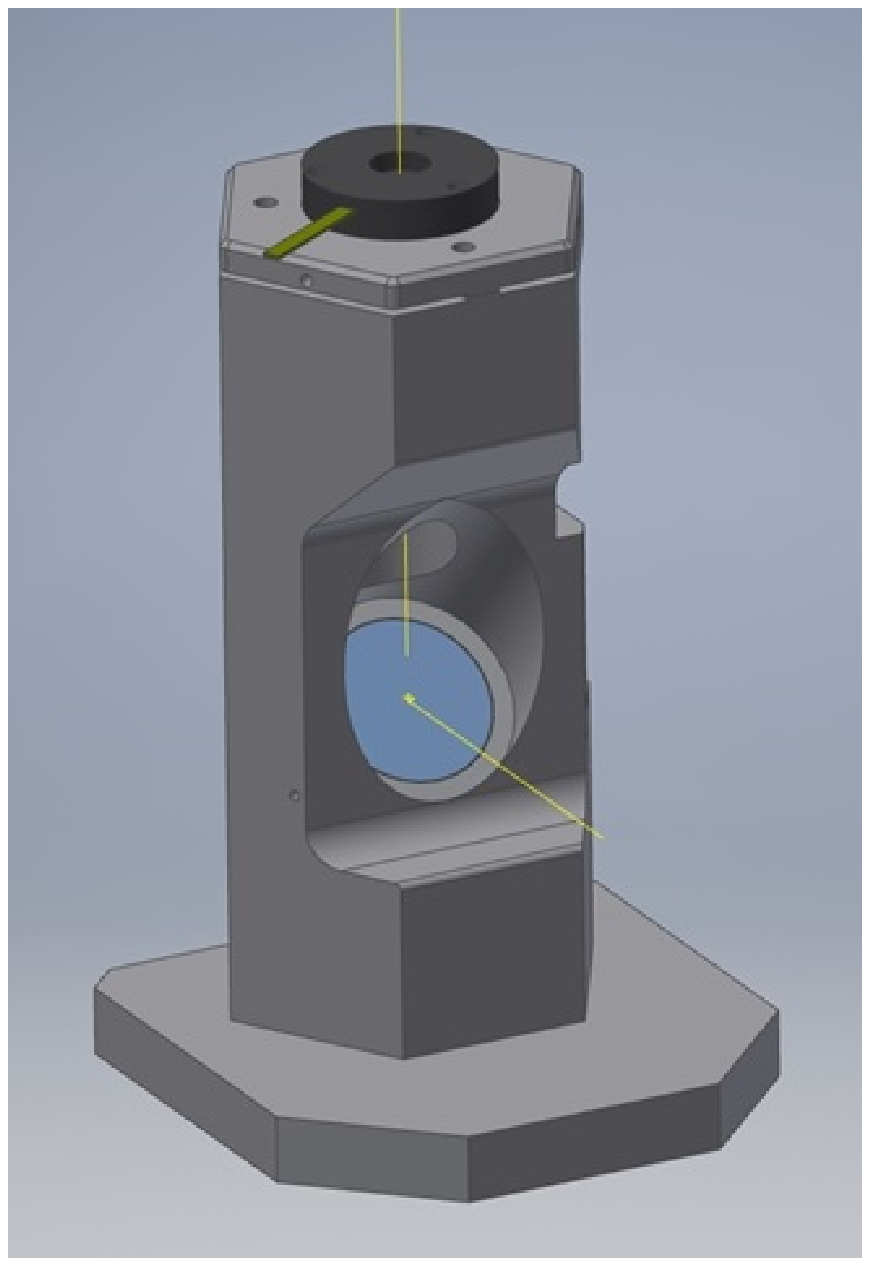}
\caption{The entrance slit of the spectrograph and the UV-VIS shutter.}
\label{fig14a} 
\end{minipage}
\hfill
\begin{minipage}[c]{0.65\linewidth}
\includegraphics[width=\linewidth]{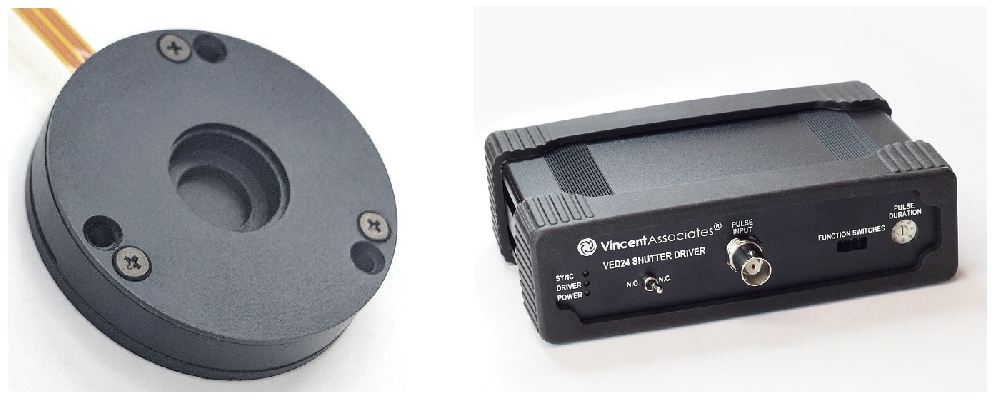}
\caption{Shutter Uniblitz ES6B and controller VED24}
\label{fig14}
\end{minipage}%
\end{figure}

The shutter is driven by the NGC through its own commercial controller; 
it is upstream from the detector unit, at the entrance slit of the spectrograph (figure~\ref{fig14a}), which allows us to use a small shutter aperture.
The selected model is an Uniblitz ES6B with its VED24 controller. (figure~\ref{fig14}). 

\textbf{Features:}
\begin{itemize}
\setlength{\itemsep}{0pt}
\item	6 mm aperture
\item	bi-stable operation
\item	extremely low-profile form-factor
\item	RoHS compliant
\item	transfer time on opening: 1.9 milliseconds
\item	total opening time: 3.7 milliseconds
\item	can be configured for the VED24 shutter driver 
\end{itemize}

\subsection{Temperature controller}
For the CCD temperature control two options can be take into account:
\begin{itemize}
\setlength{\itemsep}{0pt}
\item The CCD controller integrated control
\item A commercial temperature controller (i.e. Lakeshore 336.)
\end{itemize}

The characteristics of the Lakeshore 336 are summarized in the following list:

\begin{itemize}
\setlength{\itemsep}{0pt}
\item	operates down to 300 mK with appropriate NTC RTD sensors.
\item	four sensor inputs and four independent control outputs 
\item	two PID control loops: 100 W and 50 W into a 50 $\Omega$ or 25 $\Omega$ load
\item	auto-tuning automatically determines PID parameters 
\item   automatically switch sensor inputs using zones to allow continuous measurement and control from 300 mK to 1505 K 
\item	custom display set-up allows you to label each sensor input
\item	Ethernet, USB, and IEEE-488 interfaces 
\item	supports diode, RTD, and thermocouple temperature sensors 
\item	sensor excitation current reversal eliminates thermal EMF errors for resistance sensors 
\item	$\pm$10 V analog voltage output, alarms, and relays 
\end{itemize}


   \begin{figure} [H]
   \begin{center}
   \begin{tabular}{c} 
   \includegraphics[height=5cm]{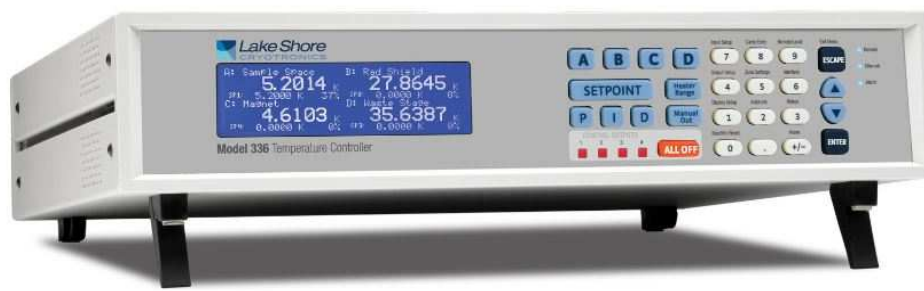}
   \end{tabular}
   \end{center}
   \caption[Shutter Temperature controller Lakeshore 335.] 
   { \label{fig15} Temperature controller Lakeshore 336.}
   \end{figure}

\section{Conclusions}
The VIS-UV detector system is part of the VIS arm of the SOXS spectrograph that will be installed at the ESO NTT telescope located at La Silla observatory. The project is going to conclude the final design phase in 2018. Afterward, the plan is to integrate the instrument in Italy and then have the first light within 2020.


 


\bibliography{VIS} 
\bibliographystyle{spiebib} 
\end{document}